\documentclass[aps,prl, twocolumn,superscriptaddress]{revtex4-2}

\usepackage{amsthm}
\usepackage{tikz}
\usepackage{amsmath}
\usepackage{amssymb}
\usepackage{color}
\usepackage{hyperref}
\usepackage{graphicx,xcolor}
\usepackage{bbold}
\usepackage{iftex}

\iftutex
\usepackage{unicode-math}
\else
\usepackage{mathrsfs}
\def\z"{}
\def\UnicodeMathSymbol#1#2#3#4{%
 \ifnum#1>"A0
   \DeclareUnicodeCharacter{\z#1}{#2}%
  \fi}
\input{unicode-math-table}

\fi

\renewcommand{\emph}[1]{\textit{#1}}

\newcommand{\ii}{\mathrm{i}} 
\newcommand{\eul}{\mathrm{e}} 
\newcommand{\diff}{\mathrm{d}} 
\newcommand{\id}{\mathbb{1}} 

\newcommand{\ket}[1]{|#1\rangle} 
\newcommand{\bra}[1]{\langle#1|} 
\newcommand{\ketbra}[2]{|#1\rangle\!\langle #2|} 
\newcommand{\tr}{\mathrm{tr}} 

\def\swap at (#1,#2){
\draw [-] (#1, #2+0.5)--(#1,#2+0.25);
\draw [-] (#1, #2+0.25)--(#1+1,#2-0.25);
\draw [-] (#1, #2-0.25)--(#1,#2-0.5);
\draw [-] (#1+1, #2+0.25)--(#1,#2-0.25);
\draw [-] (#1+1, #2+0.5)--(#1+1,#2+0.25);
\draw [-] (#1+1, #2-0.25)--(#1+1,#2-0.5);}
\def\gate at (#1,#2,#3){
\draw [-] (#1, #2+0.5)--(#1,#2-0.5);
\draw [-] (#1+1, #2+0.5)--(#1+1,#2-0.5);
\node[tensor] (1) at (#1+0.5, #2+0) {$\hspace{4mm}#3\hspace{4mm}$};
;}
\def\project at (#1,#2){
\draw [-] (#1, #2+0.5)--(#1,#2+0.25);
\draw [-] (#1+1, #2+0.5)--(#1+1,#2+0.25);
\draw [-] (#1+0.5, #2-0.25)--(#1+0.5,#2-0.5);
\draw [-] (#1, #2+0.25)--(#1+0.5,#2-0.25);
\draw [-] (#1+1, #2+0.25)--(#1+0.5,#2-0.25);
;}
\def\gateproject at (#1,#2){
\draw [-] (#1, #2+0.5)--(#1,#2-0);
\draw [-] (#1+1, #2+0.5)--(#1+1,#2-0);
\draw [-] (#1+0.5, #2+0)--(#1+0.5,#2-0.5);
\node[tensor] (1) at (#1+0.5, #2+0) {$\hspace{4mm}0\hspace{4mm}$};
\draw [-] (#1, #2+0.25)--(#1+0.5,#2-0.25);
\draw [-] (#1+1, #2+0.25)--(#1+0.5,#2-0.25);
;}

\def\gateswap at (#1,#2,#3){
\node[tensor] (1) at (#1+0.5, #2+0) {$\hspace{4mm}\hspace{4mm}#3$};
\draw [-] (#1, #2+0.5)--(#1,#2+0.25);
\draw [-] (#1, #2+0.25)--(#1+1,#2-0.25);
\draw [-] (#1, #2-0.25)--(#1,#2-0.5);
\draw [-] (#1+1, #2+0.25)--(#1,#2-0.25);
\draw [-] (#1+1, #2+0.5)--(#1+1,#2+0.25);
\draw [-] (#1+1, #2-0.25)--(#1+1,#2-0.5);}

\tikzset{every edge quotes/.style =
          { fill = white,
            sloped,
            font=\footnotesize,
}}

\definecolor{color1}{HTML}{e88f89}
\tikzstyle{tensor}=[rectangle,draw=blue!50,fill=blue!10,thick]
\tikzstyle{vector}=[rectangle,draw=color1!80,fill=color1!10,thick,rounded corners=.15cm]

\begin{document}

\title{Semi-group influence matrices for non-equilibrium quantum impurity models}

\author{Michael Sonner}
\thanks{These authors contributed equally to this work.}
\affiliation{Max Planck Institute for the Physics of Complex Systems, 01187 Dresden, Germany}

\author{Valentin Link}
\thanks{These authors contributed equally to this work.}
\affiliation{Institut f{\"u}r Theoretische Physik, Technische Universit{\"a}t Dresden, 01062, Dresden, Germany}

\author{Dmitry A. Abanin}
\affiliation{Department of Physics, Princeton University, Princeton, NJ 08544, USA}

\begin{abstract}
We introduce a framework for describing the real-time dynamics
of quantum impurity models out of equilibrium which is based on the influence
matrix approach. By replacing the dynamical map of a large fermionic quantum environment
with an effective semi-group influence matrix (SGIM) which acts on a reduced
auxiliary space, we overcome the limitations of previous proposals, achieving high
accuracy at long evolution times. This SGIM corresponds to a uniform matrix-product state representation of the influence matrix and can be obtained by an efficient algorithm presented in this paper. We benchmark this approach by computing the spectral function of the single impurity Anderson model with high resolution. Further, the spectrum of the effective dynamical map allows us to obtain relaxation rates of the impurity towards equilibrium following a quantum quench. Finally, for a quantum impurity model with on-site two-fermion loss, we compute the spectral function and confirm the emergence of Kondo physics at large loss rates.

\end{abstract}\maketitle

\paragraph{Introduction.---} 
Quantum impurity models (QIM) describe a small interacting system coupled to a large non-interacting environment. Fermionic QIMs exhibit a variety of correlated phenomena such as the Kondo effect \cite{kondo1964Resistance, hewson1997The}. These models continue to be widely studied due to their relevance to quantum transport in mesoscopic systems \cite{meir1993Lowtemperature,pustilnik2004Kondo,BullaRMP2008} and also because of their central role in the computational methods for correlated materials such as dynamical mean-field theory (DMFT)~\cite{georges1996Dynamical,kotliar2006Electronic,aoki2014Nonequilibrium}.

Experimental advances in synthetic quantum systems such as ultracold atoms enable probing real-time dynamics of QIMs~\cite{bloch2008Manybody,knap2012TimeDependent,bauer2013Realizing}. Recently, a realization of the Kondo effect in a non-equilibrium, experimentally accessible setting, where the impurity site is subject to strong dissipation, has been proposed \cite{stefanini2024Dissipative,qu2024Variational}. In such highly non-equilibrium regimes the conventional methods suited for equilibrium QIMs break down, calling for the development of new numerically exact approaches. 

In this direction, a novel class of methods for out-of-equilibrium QIMs exploits the structure of temporal correlations encoded in the environment's influence matrix (IM)~\cite{jorgensen2019Exploiting,lerose2021Influence,cygorek2022Simulation,thoenniss2023Efficient,thoenniss2023Nonequilibrium,ng2023Realtime,link2024Open,chen2024Grassmann}. The IM is a
large multi-time tensor encoding the effect of the environment on the impurity,
and, in essence, these methods aim to approximate the IM by a matrix product state
(MPS). The promise of such an approach is supported by rigorous results that
establish a polynomial scaling of complexity with evolution time~\cite{vilkoviskiy2024Bound,thoenniss2024Efficient}. While the
existing IM algorithms for fermionic QIMs
\cite{thoenniss2023Efficient,ng2023Realtime,chen2024Grassmann} are competitive
with methods such as diagrammatic Monte
Carlo~\cite{werner2009Diagrammatic,schiro2010Realtime,vanhoecke2024Diagrammatic}, hierarchical equations of motion~(HEOM)
~\cite{tanimura1989Time,jin2008Exact,tanimura2020Numerically,dan2023Efficient,thoenniss2024Efficient} and time-dependent {numerical renormalization group~(NRG)}~\cite{anders2005RealTime,nghiem2017Time,lotem2020Renormalized}, they still require significant
computational resources. 

In this paper, we demonstrate how temporal translation invariance can be
exploited to achieve a representation of the IM as a compressed uniform MPS.
This approach effectively generates a dynamical semi-group where the environment
is replaced by compressed auxiliary degrees of freedom~\cite{link2024Open}. We
introduce an efficient algorithm for constructing this semi-group influence
matrix (SGIM), achieving accurate and controlled computations of non-equilibrium
QIM dynamics for long evolution times. In particular, as we show below, the SGIM
method allows for computing a spectral function of the single-impurity Anderson
model with high resolution across all frequencies. Furthermore, we study the dynamical
formation of Kondo correlations following a quantum quench, and, finally,
investigate the dissipative Kondo
effect \cite{stefanini2024Dissipative,qu2024Variational} using this numerically
exact method.

\paragraph{Method.---}
We consider a QIM where the impurity ($\mathrm{I}$) consists of fermionic modes labeled by an
index $\sigma$. Each of these modes is coupled to an independent Gaussian
fermionic environment (E) via the Hamiltonian
\begin{equation}\label{eq:Hamiltonian}
    H_\mathrm{IE}^\sigma=\sum_k (v_{\sigma k} d^\dagger_\sigma f_{\sigma k}+v_{\sigma k}^* f_{\sigma k}^\dagger d_\sigma)+\sum_k\varepsilon_{\sigma k} f_{\sigma k}^\dagger f_{\sigma k}.
\end{equation}
Here, $d_\sigma$ and $f_{\sigma k}$ are the annihilation operators for the
fermions in the impurity and the environment, respectively. Each environment is
fully characterized by its spectral density $J_\sigma(ω) = \sum_k |v_{\sigma
k}|^2\delta(\omega - \varepsilon_{\sigma k}) $, inverse temperature $\beta$ and chemical
potential $\mu$ \cite{devega2015How}.

\begin{figure}
    \includegraphics[scale=0.075]{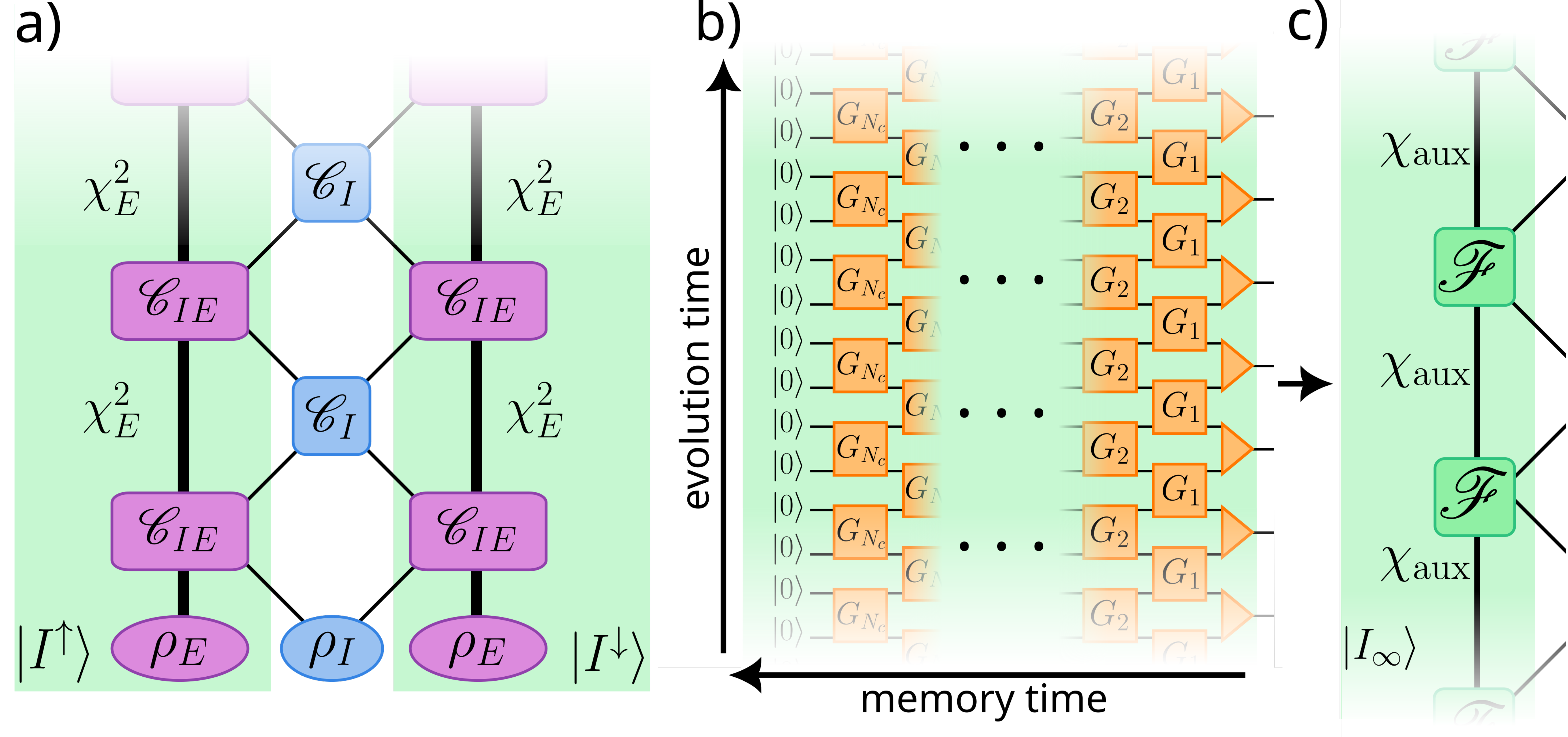}
    \caption{(a) Real time evolution of a QIM with two baths $\sigma=\uparrow,\downarrow$ expressed as tensor contraction
    corresponding to Eq.~\eqref{eq:dmevo}. (b) Tensor network representation
    of an infinite Gaussian IM corresponding to Eq.~\eqref{eq:F_inf_def},
    decomposed into nearest neighbor gates. (c) Semi-group influence matrix as
    obtained from contracting (b) using iTEBD.}
    \label{fig:fig1}
\end{figure}

For a small
time-step $\delta t$, the dynamical map or \emph{quantum channel} of the QIM dynamics can be expressed within second order Trotter approximation
as
\begin{align}
    \mathscr{C}_\mathrm{QIM}[δt] = \mathscr{C}_\mathrm{I}[δt/2] \left(∏_{σ} \mathscr{C}^σ_\mathrm{IE}[δt]\right)\mathscr{C}_\mathrm{I}[δt/2] + O(δt^3), \label{eq:qimchannel}
\end{align} 
where $\mathscr{C}_\mathrm{I}[δt/2]$ is a channel for a local, interacting evolution on the impurity (half step $δt/2$) and the evolution $\mathscr{C}^\sigma_\mathrm{IE}[δt]=\eul^{-\ii H_\mathrm{IE}^\sigma\delta t}\otimes \eul^{\ii H_\mathrm{IE}^\sigma\delta t}$
describes the coupling with the environment via Hamiltonian~\eqref{eq:Hamiltonian}. The channel
\eqref{eq:qimchannel} can be used to express the evolution of the impurity state for $N$ time steps 
(c.f.~Fig.~\ref{fig:fig1}a)
\begin{align}
    ρ_\mathrm{I}(Nδt) = \tr_\mathrm{E}\left[\left(\mathscr{C}_\mathrm{QIM}[δt]\right)^N ρ_\mathrm{QIM}(0)\right] \label{eq:dmevo}
\end{align}
where $\tr_\mathrm{E}$ denotes the trace over all environment modes.
We can collect all terms in Eq.~\eqref{eq:dmevo} that correspond to a single
environment in one object (green shaded area in Fig.~\ref{fig:fig1}a). This
defines the \emph{influence matrix}~(IM) \cite{lerose2021Influence},
\begin{align}\label{eq:IM_def}
    \ket{I_N^\sigma} = \tr_\mathrm{E} \left[\left(\mathscr{C}_\mathrm{IE}^{σ}[\delta t]\right)^{⊗_\mathrm{I} N}\rho_\mathrm{E}^\sigma\right],
\end{align}
where $⊗_\mathrm{I}$ denotes the tensor product over the impurity degrees
of freedom, while the environment degrees of freedom are contracted. In the
context of characterizing non-Markovian dynamics, a similar object is known as
\emph{process tensor} \cite{pollock2018Non}. Time evolution as in
Eq.~\eqref{eq:dmevo} can be expressed in terms of overlaps of IMs
\cite{thoenniss2023Efficient,thoenniss2023Nonequilibrium}, which can be computed
efficiently if the IM is parametrized as a matrix product state~(MPS) with low
bond dimension. However, for any given MPS-IM, the maximal number of time step
which can be computed is limited by $N$.

On the other hand, using Eq.~\eqref{eq:dmevo} directly for numerical
computations is impractical because the channels $\mathscr{C}^σ_\mathrm{IE}$
act on the large environment density matrix space. The central idea of our
approach is to find an efficient algorithm to replace 
these channels by a different linear map $\mathscr{F}^σ$ that
faithfully reproduces the local impurity dynamics when used in
Eq.~\eqref{eq:dmevo}, but acts on a much smaller \emph{auxiliary space}. This
linear map defines a \emph{semi-group influence matrix}~(SGIM) and is a representation of the effective dynamical {semi-group} of the
environment time-evolution \cite{kossakowski1972quantum}. The accuracy of this representation as well as the numerical cost
is controlled by the dimension of the auxiliary space $\chi_\mathrm{aux}$. 
By substituting $\mathscr{F}^σ$ into Eq.~\eqref{eq:IM_def}, we can understand it
as a uniform MPS representation of an IM with infinite number of time steps (Fig.~\ref{fig:fig1}c).




Previous works \cite{thoenniss2023Efficient, ng2023Realtime,
thoenniss2023Nonequilibrium} showed that for Gaussian fermionic environments in QIMs, the IM can, in a vectorized form, be expressed as a
Gaussian many-body state
\begin{equation}\label{eq:F_def}
    \ket{I_N^\sigma}=\prod_{n=1}^N\prod_{m=1}^{n}\exp\left(\Vec{c}_n^\dagger \cdot G_{n,m}^\sigma\Vec{c}_m^\dagger\right)\ket{0},
\end{equation}
where $G_{n,m}$ are
$4\times 4$ correlation matrices, and $\vec{c}_n$ are fermionic annihilation operators with
four fermionic modes per time-step $n$ (see \cite{supp_mat}). Various algorithms
exist to directly generate a finite MPS representation of states in the form
\eqref{eq:F_def} \cite{petrica2021Finite,jin2022Matrix,fishman2015Compression,ng2023Realtime},
resulting in an efficient algorithm for short time dynamics of QIMs
\cite{thoenniss2023Efficient,ng2023Realtime}.

Looking at the correlation matrices $G^\sigma_{m,n}$ closely, we
note that in the bulk, for $1\ll n,m\ll N$, the temporal correlations become
effectively translationally invariant $G^\sigma_{n,m}=G^\sigma_{n-m}$.
Furthermore, the correlations of continuous environments decay at large time differences.
This allows us to neglect correlations for time differences larger than some
bath dependent effective memory time $N_c$. We can formally express an ``infinite
IM" as
\begin{equation}\label{eq:F_inf_def}
    \ket{I_\infty^\sigma}=\prod_{n=-\infty}^{+\infty}\prod_{k=0}^{N_c}\exp\left(\Vec{c}_n^\dagger \cdot G_{k}^\sigma\Vec{c}_{n-k}^\dagger\right)\ket{0}.
\end{equation}
This state is translationally invariant and can thus be efficiently
represented as a uniform MPS, allowing us to find an effective SGIM $\mathcal{F}^\sigma$. Note that, even
though the SGIM is time translation invariant, it still can be
used to compute non-stationary impurity dynamics such as local quenches by using appropriate time-dependent evolution on the impurity.
To construct a SGIM, we first
derive an exact tensor network representation of Eq.~\eqref{eq:F_inf_def},
which is depicted in Fig.~\ref{fig:fig1}b. Briefly, each network layer introduces correlations $G_{k}^\sigma$ corresponding to a given memory time step $k$ in Eq.~\eqref{eq:F_inf_def}, 
starting from the memory time cutoff $k=N_c$ and going to $k=0$. Details
on this construction are provided in the supplement \cite{supp_mat}. Structurally,
our result resembles the network used in the TEMPO method for bosonic Gaussian
environments \cite{strathearn2018Efficient, link2024Open}. 


Contracting the tensor network (see Fig.~\ref{fig:fig1}c) to uniform MPS using conventional
infinite time-evolving block decimation~(iTEBD) \cite{vidal2007Classical} yields
a SGIM. 
Truncation with a fixed SVD (singular value decompositon) truncation tolerance is highly efficient because entanglement, and hence the MPS bond dimension, will
build up only towards the end of the contraction. This allows us to set the
memory time cutoff $N_c$ so large that all correlations beyond the cutoff fall
below the truncation tolerance, eliminating the memory time cutoff as a
parameter which needs to be monitored for convergence. 


Compared to finite-MPS approaches, the SGIM method allows for significantly
longer evolution times with lower demands in computation time and memory. The
construction of the SGIM is dominated by singular value decompositions in the
last few layers of the network and is independent of the number of time steps. 
This is in contrast to the construction of a finite MPS for $N$ time steps
which requires, depending on the approach, $O(N)$ \cite{thoenniss2023Efficient} or $O(N^2)$ \cite{ng2023Realtime} SVD computations with large bond dimension. An alternative method to construct an infinite MPS-IM within an auxiliary mode expansion \cite{guo2024Infinite} relies on iDMRG, which is likewise computationally very demanding.
With our algorithm, IM simulations are no longer limited by the compression
step. 

Owing to the semi-group structure, the SGIM is very well-suited for computing
dynamical properties of QIMs in the stationary regime. For this we first compute the stationary state corresponding to
the leading eigenvector of the
auxiliary dynamical map, either by Krylov methods \cite{lehoucq1998ARPACK} or by
simply running the time evolution until convergence (power iteration). The stationary state is then used as initial condition to compute the real-time Green's function
from time evolution, which can be Fourier transformed to obtain the spectral
function. Since our method captures the full continuum bath, no postprocessing
such as spectral broadening \cite{lee2016Adaptive} is required.

\paragraph{Equilibrium benchmark.---}
To benchmark our method we first consider the single impurity Anderson
model~(SIAM) \cite{anderson1961Localized} where spin-1/2 fermions
($\sigma=\uparrow,\downarrow$) interact on a single site with a Hubbard
interaction in an otherwise non-interacting system. The impurity Hamiltonian for
this model is given by
\begin{align}\label{eq:SIAM}
    H_\mathrm{I}=U n_\uparrow n_\downarrow - \frac{U}{2} n_\uparrow - \frac{U}{2} n_\downarrow.
\end{align} 
Equilibrium properties of the SIAM can be computed with a variety of established
approaches \cite{hewson1997The,kotliar2006Electronic,gull2011Continuous}, so that excellent reference data is available.

\begin{figure}
    \includegraphics[width=\columnwidth]{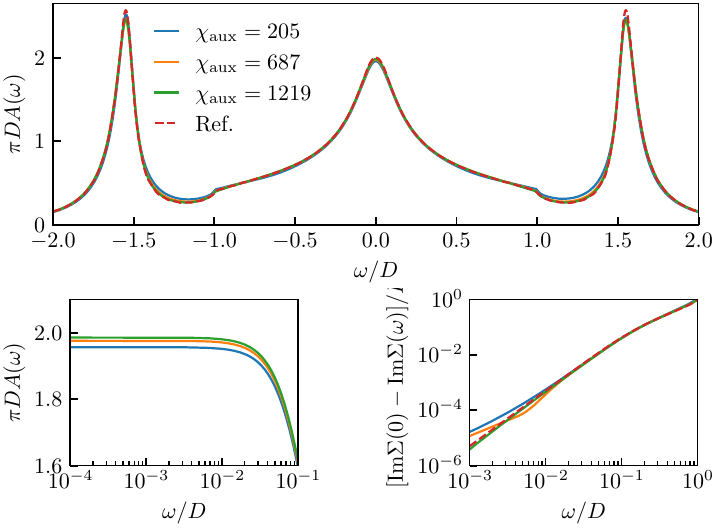}
    \caption{ Upper: Spectral function $A(\omega)$ of the SIAM \eqref{eq:SIAM}
    with semicircular bath
    $J_{\uparrow/\downarrow}(\omega)={D}/(2\pi)\sqrt{1-\omega^2/D^2}$,
    $\mu=0,\,\beta=100/D$ and $U=2D$. Lower: Left: spectral function at low
    frequencies. Friedel sum rule requires $\pi DA(0)=2$ at $\beta=\infty$.
    Right: Imaginary part of the self energy $\Sigma(\omega)$ at low
    frequencies (reference data from Ref.~\cite{cao2024Dynamical}). We used a
    Trotter time-step $\delta t=0.05/D$.}
    \label{fig:semicircle_sf}
\end{figure}

In Fig.~\ref{fig:semicircle_sf} we display the spectral function for a benchmark
problem considered in Refs.~\cite{grundner2024Complex,cao2024Dynamical}. The spectral function shows the Kondo resonance at low frequencies, Hubbard resonances at high frequencies, as well as the sharp band-edges. The agreement with the reference is already very good at moderate auxiliary space dimensions $\chi_\mathrm{aux}$. Crucially, using higher accuracy representations we obtain excellent low-frequency resolution for the spectral
function, recovering almost the entire spectral weight at
$\omega=0$, with an error below 1\% for the Friedel sum rule $\pi DA(0)=2$ \cite{luttinger1961Analytic}. The expected Fermi-liquid behavior of the self energy is also
fulfilled up to small frequencies, with consistent improvement when the
auxiliary space dimension is increased. 

\begin{figure}
    \includegraphics[width=\columnwidth]{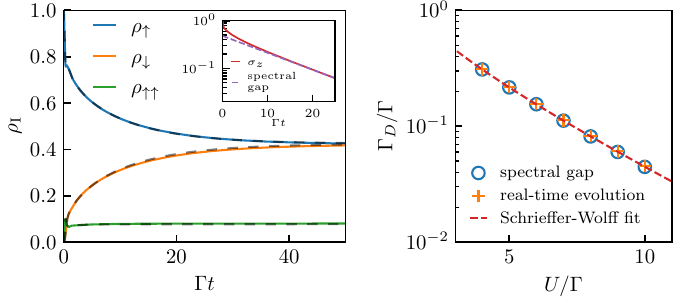}
    \caption{SIAM dynamics for a flatband bath
    $J_{\uparrow/\downarrow}(\omega)={2\Gamma}/[2\pi(1+\eul^{\nu(\omega-\omega_c)})(1+\eul^{-\nu(\omega+\omega_c)})]
    $ with $\nu=\omega_c=10\Gamma$, $\beta=100/\Gamma$, $\mu=0$. Left: Quench
    from a polarized impurity with $U=8\Gamma$. We achieve convergence with
    $\delta t=0.025/\Gamma,\,\chi_\mathrm{aux}=305$ (dashed:
    $\chi_\mathrm{aux}=154$). Insert: Asymptotic exponential decay of
    $\sigma_z=\rho_{\uparrow}-\rho_{\downarrow}$ compared to the prediction from
    the   spectral gap.  Right: Spin relaxation rate $\Gamma_D$ (computed via
    the spectral gap and real time evolution) for different interaction
    strengths $U$, and fit to the Schrieffer Wolff formula
    \cite{hewson1997The,cao2024Dynamical}.} \label{fig:hubbard_kondo_quench}
\end{figure}

\paragraph{Quench and Kondo physics.---} 
We now turn to a non-equilibrium situation where dynamics is triggered by a
quench from an initially polarized impurity. In
Fig.~\ref{fig:hubbard_kondo_quench} we consider a flatband environment, particularly well-suited to study Kondo physics. Real-time dynamics of this model was
considered previously in Refs.~\cite{cohen2015Taming, thoenniss2023Efficient} for limited evolution times. 
With our approach, we are now able to describe the spin relaxation for long times, while retaining high accuracy for the
fast initial quench dynamics. To characterize the asymptotic decay, we can make
use of semi-group formalism by extracting dynamical properties from the
spectrum of our effective dynamical map. For example,
we can extract the longest time scale of the problem directly as the logarithm
of the next-to-leading eigenvalue. In the SIAM this time scale is the asymptotic
spin relaxation rate $\Gamma_D$, which corresponds to the Kondo scale
\cite{nuss2015Nonequilibrium,lechtenberg2014Spatial}.
The next-to-leading eigenvalue, or spectral gap, can be computed efficiently
using Krylov methods \cite{lehoucq1998ARPACK}. As displayed in
Fig.~\ref{fig:hubbard_kondo_quench} (insert), the extracted rate indeed predicts the slow
spin relaxation at long times. We reproduce the exponential dependence of the relaxation rate on the interaction strength $U$, finding an excellent fit with the functional form $\Gamma_D(U)={\sqrt{a/U}}\eul^{-b U}$ expected for the Kondo scale in the Schrieffer-Wolff limit \cite{hewson1997The,cao2024Dynamical}.

\begin{figure}
    \includegraphics[width=\columnwidth]{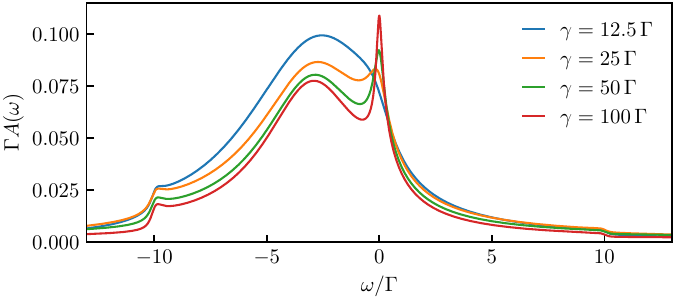}
    \caption{Spectral function for the dissipative SIAM with the same bath used
    in Fig.~\ref{fig:hubbard_kondo_quench} and onsite energy
    $\varepsilon_d=-3\Gamma$. As the loss rate $\gamma$ is increased, a sharp
    Kondo peak appears at $\mu=0$. For these computations we used a Trotter time-step of $\delta t=0.002/\Gamma$ and an auxiliary dimension of
    $χ_\mathrm{aux}=239$.}\label{fig:diss_kondo_sf}
\end{figure}

\paragraph{Dissipative impurity model.---}
A key advantage of IM approaches is their ability to seamlessly handle
non-unitary impurity
dynamics such as Markovian dissipation
\cite{sonner2021Influence,mi2022Noiseresilient} without incurring additional
numerical overhead. 
External dissipation creates a genuine non-equilibrium situation, opening new
possibilities to engineer and explore unconventional quantum states
\cite{sieberer2016Keldysh}. For instance, recent works proposed a way to realize
Kondo physics with strong local two-fermion losses
\cite{stefanini2024Dissipative,qu2024Variational}, with local dynamics generated
by the Lindbladian
\begin{align}
    \mathscr{L}_\mathrm{I}\rho &= -\ii \varepsilon_d [n_\uparrow + n_\downarrow,\rho]+\gamma L\rho L^\dagger - \frac{\gamma}{2}\{L^\dagger L,\rho  \}
\end{align}
where $L=d_\uparrow d_\downarrow$. The corresponding local channel is given by
$\mathscr{C}_\mathrm{I}[δt/2]= e^{\mathscr{L}_\mathrm{I} δt/2}$.
In the limit $\gamma\rightarrow\infty$ a double occupancy of the impurity is
completely suppressed and the model reduces to the Anderson impurity model with
infinite repulsion, a model known to exhibit Kondo physics. For finite $\gamma$,
the emergence of the Kondo effect has been explored with adiabatic elimination
\cite{stefanini2024Dissipative} and a multi-Gaussian variational ansatz
\cite{qu2024Variational}.
Here, we compute the \emph{exact} spectral function of
the model for different values of $\gamma$ with a flatband bath, shown in
Fig.~\ref{fig:diss_kondo_sf}. Crucially, our method can cover the full crossover
at all frequencies without restrictive approximations. At small $\gamma$
the spectral function is close to the noninteracting case, with a single peak centered at the
impurity energy $\varepsilon_d$, broadened by the interaction with the bath. As
the loss rate is increased, a sharp Kondo peak
develops at $\mu=0$, corresponding to the slow spin relaxation also
present in the unitary SIAM (Fig.~\ref{fig:hubbard_kondo_quench}).

\paragraph{Conclusions.---} 
The SGIM construction introduced here is an efficient approach to studying
out-of-equilibrium QIMs, giving access to long evolution times with low
computational demands. The underlying \emph{temporal} MPS representation
provides a significantly more effective compression of the environment compared
to direct time evolution of QIMs via chain mappings~\cite{chin2010Exact} in
combination with regular spatial
MPS~\cite{ganahl2015Efficient,schroder2016Simulating,wauters2024Simulations,manaparambil2024Nonequilibrium},
which requires large bond dimensions to achieve convergence at low
frequencies~\cite{cao2024Dynamical}. In the future, our method can be applied to
a wide range of problems. In particular, this includes transport phenomena in
strongly correlated systems, where real-time dynamics are required to compute
non-equilibrium steady states ~\cite{manaparambil2024Nonequilibrium}. Further,
the IM framework can be used directly for driven impurities beyond the
high-frequency limit~\cite{kahlert2024Simulating}, as well as for feedback and
optimal control problems~\cite{ortega2024Unifying} and as impurity solver for
non-equilibrium DMFT \cite{aoki2014Nonequilibrium,nayak2025Steady}. Finally,
the efficient compressed representation of environments could offer a favorable
scalability for numerical simulations of multi-orbital impurity problems, which
remains a central challenge in multi-orbital DMFT.

\paragraph{Acknowledgments.---}
V.L. acknowledges discussions with Hong-Hao Tu. M.S. acknowledges discussions
with Matan Lotem, Julian Thoenniss and Alessio Lerose. Computations were
performed on the HPC system Ada at the Max Planck Computing and Data Facility
and on the internal HPC cluster of the Max Planck Institute for the Physics of
Complex Systems.

\bibliography{bib.bib}
\clearpage
\onecolumngrid

\setcounter{equation}{0}
\renewcommand{\theequation}{S\arabic{equation}}
\renewcommand{\thesection}{SM\arabic{section}}
\renewcommand{\thefigure}{S\arabic{figure}}

\begin{center}
{\large \bf Supplemental Material}
\end{center}

\section{Correlation Matrix}

In this section we provide details on the correlation matrix appearing in the influence matrix (IM) for fermionic impurities. We consider a single fermion species only and do not write explicitly the impurity mode label $\sigma$. To define the vectorized IM we introduce two fermionic modes (``in" and ``out") at each time step on the discretized Keldysh contour. Respectively, the IM is a state of 4$N$ fermionic modes, two in/out modes at each time step $(n\leq N)$ on both the forward ($+$) and backward ($-$) branches of the contour. We combine the
corresponding annihilation operators in a vector $\Vec{c}_n=(c_{n}^{\mathrm{in},+},c_{n}^{\mathrm{in},-},c_{n}^{\mathrm{out},+},c_{n}^{\mathrm{out},-})$.
Up to normalization, the IM can then be expressed in terms of the Gaussian fermionic many-body state Eq.~\eqref{eq:F_def} (main text).

The exact correlation matrix $G_{n,m}$ of the IM has been derived in Ref.~\cite{ng2023Realtime}. It reads for $n>m$
\begin{equation}
G_{n,m}=\begin{pmatrix}
    0 & (G_-(n+1,m+1))^* & G_<^{++}(n+1,m) & 0\\
    -G_-(n+1,m+1) & 0 & 0 & -(G_<^{++}(n+1,m))^*\\
    G_>^{++}(n,m+1) & 0 & 0 & (G_+(n,m))^*\\
    0 & -(G_>^{++}(n,m+1))^* & -G_+(n,m) &0
    \end{pmatrix},
\end{equation}
and for $n=m$
\begin{equation}
G_{n,n}=\begin{pmatrix}
    0 & 0 & 0 & 0\\
    -G_-(n+1,n+1) & 0 & 0 & 0\\
    F-G_<^{++}(n+1,n) & 0 & 0 & 0\\
    0 & -F^*+(G_<^{++}(n+1,n))^* & -G_+(n,n) &0
    \end{pmatrix}.
\end{equation}
The functions $G_>^{++}, G_<^{++}, G_+, G_-$ and $F$ are provided in the supplement of Ref.~\cite{ng2023Realtime}. Note that these are in general not translationally invariant $G_{k+m,m}\neq G_{k+n,n}$. However, for continuous environments, they become translationally invariant at large times (and also at $\mathcal{O}(\delta t^2)$), which yields a translationally invariant IM in the infinite time limit. We can evaluate the correlation matrix exactly by solving the Gaussian (single particle) dynamics. For this we define the unitary matrix
\begin{equation}
\exp\left(-\ii \delta t  \begin{pmatrix}
      0 & (v_k)^T \\
      (v_k^*) & \mathrm{diagm}(\varepsilon_k)
    \end{pmatrix}\right)\equiv 
\begin{pmatrix}
    F & (x_k)^T\\
    (y_k) & U
\end{pmatrix}
\end{equation}
as well as the Boltzmann weights
\begin{equation}
\rho_B=\mathrm{diagm}\!\left(\eul^{-\beta(\varepsilon_k-\mu)} \right).
\end{equation}
Here, $\varepsilon_k$ and $v_k$ are the bath parameters from Eq.~\eqref{eq:Hamiltonian} (main text).
Using these definitions we can express the relevant functions explicitly:
\begin{align}\label{eq:G_mat1}
G_>^{++}(n,m)&=\boldsymbol{x}^TU^{n-N}\frac{1}{\id+U^N\rho_B(U^N)^\dag}U^{N-m}\boldsymbol{y}\\
G_<^{++}(n,m)&=\boldsymbol{x}^TU^{m-N}\left[\id-\frac{1}{\id+U^N\rho_B(U^N)^\dag}\right]U^{N-n}\boldsymbol{y}\\
G_-(n,m)&=\boldsymbol{y}^\dagger (U^{N-n})^\dagger \frac{1}{\id+U^N\rho_B(U^N)^\dag} U^{N-m}\boldsymbol{y}\\
G_+(n,m)&=\boldsymbol{x}^T U^{m-N}\left[\id-\frac{1}{\id+U^N\rho_B(U^N)^\dag}\right](U^{n-N})^\dag\boldsymbol{x}^*\label{eq:G_mat4}
\end{align}
Note that these expressions can be evaluated directly using a discretization of the continuous environment in terms of $M$ modes ($k=1,...,M$), with only $\mathcal{O}(M^3)$ numerical effort. 

To establish a connection to the more familiar continuum path integral expressions, we also provide the leading order in $\delta t$ in the following:
\begin{align}
        G_{>}^{++}(m+k,m+1)&=-\delta t^2g_>(k\delta t)+\mathcal{O}(\delta t^3)\\
        G_<^{++}(m+k,m+1)&=\delta t^2(g_<(k\delta t))^*+\mathcal{O}(\delta t^3) \\
        G_{-}(m+k,m)&=\delta t^2 g_>(k\delta t)+\mathcal{O}(\delta t^3)\\
        G_{+}(m+k+1,m)&=-\delta t^2(g_<(k\delta t))^*+\mathcal{O}(\delta t^3)\\
        F-1&=\frac{\delta t^2}{2}(g_<(0)-g_>(0))+\mathcal{O}(\delta t^3)
\end{align}
The temporal greens functions $g_\lessgtr(t)$, are directly related to the bath spectral density
\begin{equation}
    g_>(t)=\int\diff\omega  J(\omega)(1-n_F(\omega))\eul^{-\ii\omega t},\qquad g_<(t)=\int\diff\omega  J(\omega)(-n_F(\omega))\eul^{-\ii\omega t}.
\end{equation}
Using these ``continuum" approximations for the correlation matrix will cause additional Trotter errors which we avoid through computing the exact expressions \eqref{eq:G_mat1}-\eqref{eq:G_mat4}.

\section{Network Construction}\label{app:network_construction}

In this section we construct an exact tensor network representation of the fermionic IM that can be contracted efficiently to a uniform matrix product state (MPS) via infinite time-evolving block decimation (iTEBD). We consider here exclusively the infinite time limit, although the same scheme could also be used to obtain a finite tensor network representation for the finite-$N$ influence matrix. Moreover, we omit writing the impurity mode index $\sigma$ in the following. We first introduce for each fermion species $c_{n}^\lambda$ ($\lambda=+/-,\mathrm{in/out}$) in the IM two independent auxiliary fermions $a_{n}^\lambda$ and $b_{n}^\lambda$. Then we define the following linear and parity conserving mapping between the Hilbert space of the $a_{n}^\lambda, b_{n}^\lambda$ modes and the original Hilbert space of the physical $c_{n}^\lambda$ mode
\begin{equation}
    P_n^\lambda=\ketbra{0}{00}+\ket{1}\big(\bra{01}+\bra{10}\big).
\end{equation}
At the level of a Gaussian paired state, such as the IM Eq.~\eqref{eq:F_inf_def} (main text), this projection allows the formal replacements $\vec{c}_n^\dagger\rightarrow \vec{a}_n^\dagger$ or $\vec{c}_n^\dagger\rightarrow \vec{b}_n^\dagger$. 
In particular, the IM can be expressed in the following way
\begin{equation}\label{eq:F_inf_proj}
    \ket{I_\infty}=\prod_{n=-\infty}^{+\infty}P_n\prod_{k=0}^{N_c}\exp\left(\vec{a}_n^\dagger \cdot G_{k}\vec{b}_{n-k}^\dagger\right)\ket{0},
\end{equation}
where $\ket{0}$ now denotes the $a,b$ vacuum. We also define two sets of fermionic swap operators via
\begin{equation}
    S_n^{(1)} \vec{a}^\sigma_n = \vec{b}^\sigma_n S_n^{(1)},\qquad S_n^{(2)} \vec{b}_n = \vec{a}_{n+1} S_n^{(2)}.
\end{equation}
The IM can now be constructed from applying local nearest-neighbor gates as follows: \\
In the first (odd) step we set
\begin{equation}
\begin{split}
    \ket{\psi_1}&=\prod_{n=-\infty}^{+\infty} S_n^{(1)} \exp\left(\Vec{a}_n^\dagger \cdot G_{N_c}\Vec{b}_{n}^\dagger\right)\ket{0}=\prod_{n=-\infty}^{+\infty}\exp\left(\Vec{b}_n^\dagger \cdot G_{N_c}\Vec{a}_{n}^\dagger\right)\ket{0},
\end{split}
\end{equation}
using that $\ket{0}$ is swap invariant. In the second (even) step we apply
\begin{equation}
\begin{split}
    &\ket{\psi_2}=\prod_{n=-\infty}^{+\infty} S_n^{(2)}\exp\left(\Vec{b}_{n}^\dagger \cdot G_{N_c-1}\Vec{a}_{n+1}^\dagger\right)\ket{\psi_1}=\prod_{n=-\infty}^{+\infty} \exp\left(\Vec{a}_{n+1}^\dagger \cdot G_{N_c-1}\Vec{b}_{n}^\dagger\right) \exp\left(\Vec{a}_{n+1}^\dagger \cdot G_{N_c}\Vec{b}_{n-1}^\dagger\right)\ket{0}.
\end{split}
\end{equation}
We continue this procedure with alternating odd and even steps:
\begin{equation}
\begin{split}
    \ket{\psi_3}&=\prod_{n=-\infty}^{+\infty} S_n^{(1)} \exp\left(\Vec{a}_n^\dagger\cdot  G_{N_c-2}\Vec{b}_{n}^\dagger\right)\ket{\psi_2}
    \\&=\prod_{n=-\infty}^{+\infty}\exp\left(\Vec{b}_n^\dagger \cdot G_{N_c-2}\Vec{a}_{n}^\dagger\right)\exp\left(\Vec{b}_{n+1}^\dagger \cdot G_{N_c-1}\Vec{a}_{n}^\dagger\right) \exp\left(\Vec{b}_{n+1}^\dagger \cdot G_{N_c}\Vec{a}_{n-1}^\dagger\right)\ket{0}
\end{split}
\end{equation}
\begin{equation}
\begin{split}
       \ket{\psi_4}&=\prod_{n=-\infty}^{+\infty} S_n^{(2)}\exp\left(\Vec{b}_{n}^\dagger \cdot G_{N_c-3}\Vec{a}_{n+1}^\dagger\right)\ket{\psi_2}\\
       &=\prod_{n=-\infty}^{+\infty} \exp\left(\Vec{a}_{n+1}^\dagger \cdot G_{N_c-3}\Vec{b}_{n}^\dagger\right)\exp\left(\Vec{a}_{n+1}^\dagger \cdot G_{N_c-2}\Vec{b}_{n-1}^\dagger\right)\exp\left(\Vec{a}_{n+2}^\dagger \cdot G_{N_c-1}\Vec{b}_{n-1}^\dagger\right) \exp\left(\Vec{a}_{n+2}^\dagger \cdot G_{N_c}\Vec{b}_{n-2}^\dagger\right)\ket{0}\\
       &=\prod_{n=-\infty}^{+\infty} \exp\left(\Vec{a}_{n}^\dagger \cdot G_{N_c-3}\Vec{b}_{n-1}^\dagger\right)\exp\left(\Vec{a}_{n}^\dagger \cdot G_{N_c-2}\Vec{b}_{n-2}^\dagger\right)\exp\left(\Vec{a}_{n}^\dagger \cdot G_{N_c-1}\Vec{b}_{n-3}^\dagger\right) \exp\left(\Vec{a}_{n}^\dagger \cdot G_{N_c}\Vec{b}_{n-4}^\dagger\right)\ket{0}
\end{split}
\end{equation}
Assuming even $N_c$, we obtain after $N_c$ steps
\begin{equation}
    \ket{\psi_{N_c}}=\prod_{n=-\infty}^{+\infty}\prod_{k=1}^{N_c}\exp\left(\vec{a}_n^\dagger \cdot G_k \vec{b}_{n-k}^\dagger\right)\ket{0}.
\end{equation}
In the final evolution step we apply the projection $P$ instead of the swap
\begin{equation}
    \ket{I_\infty}=\prod_{n=-\infty}^{+\infty}P_n\exp\left(\vec{a}_n^\dagger \cdot G_0 \vec{b}_{n}^\dagger\right)\ket{\psi_{N_c}},
\end{equation}
recovering the exact IM via Eq.~\eqref{eq:F_inf_proj}. Since the network consists only of local gates between neighboring $\vec{a}$ and $\vec{b}$ modes, it can be contracted directly via iTEBD. Note that, at the beginning of the contraction, the value of the correlation matrix $G_k$ is small and the state $\ket{\psi_{N_c-k}}$ is almost swap invariant. The action of the nearest-neighbor gates is therefore weakly entangling and the required bond dimension of the MPS grows slowly.

We can write the network pictorially by introducing the notations
\begin{equation}
\begin{tikzpicture}
\node (e) at (0,0) {$\eul^{\Vec{a}^\dagger\cdot G_k \Vec{b}^\dagger}=$};
\gate at (1.1,0,k);
\end{tikzpicture},\hspace{1cm}
\begin{tikzpicture}
\node (e) at (0,0) {$S=$};
\swap at (0.6,0);
\end{tikzpicture}\,,\hspace{1cm}
\begin{tikzpicture}
\node (e) at (0,0) {$P=$};
\project at (0.7,0);
\end{tikzpicture},
\end{equation}
where the gates are to be understood as acting on states of neighboring $\vec{a}, \vec{b}$ modes via
\begin{equation}
\begin{tikzpicture}
\node (e) at (0,0.3) {$\eul^{\Vec{a}^\dagger\cdot G_k \Vec{b}^\dagger}(\ket{\psi_a}\otimes\ket{\psi_b})=$};
\gate at (2.3,0,k);
\node[vector] at (2.3,0.8) {$\!\ket{\psi_a}\!$};
\node[vector] at (3.3,0.8) {$\!\ket{\psi_b}\!$};
\end{tikzpicture}.
\end{equation}
For a more compact notation we further define the combined gates
\begin{equation}
\begin{tikzpicture}
    \swap at (0,0);
    \gate at (0,1,k) ;
    \node at (1.5,0.5) {$=$};
    \gateswap at (2,0.5,k);
\end{tikzpicture},\hspace{2cm}
\begin{tikzpicture}
    \project at (0,0);
    \gate at (0,1,0) ;
    \node at (1.5,0.5) {$=$};
    \gateproject at (2,0.5);
\end{tikzpicture}.
\end{equation}
Then the infinite IM is given by the following network ($N_c=3$ in this example):
\begin{equation}
\begin{tikzpicture}
\foreach \j in {1,...,10} {
    \node[vector] (v) at (\j , 0) {$\ket{0}$};
    \draw[-] (v)-- (\j,-0.5);
};
\foreach \j in {0,...,4} {
    \gateswap at (1+\j+\j, -2, 2);
    \gateproject at (1+\j+\j, -4);
};
\foreach \j in {0,...,3} {
    \gateswap at (2+\j+\j, -1, 3);
    \gateswap at (2+\j+\j, -3, 1);
};
\node at (1,-1) {$\cdots$};
\node at (1,-3) {$\cdots$};
\node at (10,-1) {$\cdots$};
\node at (10,-3) {$\cdots$};
\node at (1.5,-4.8) {$\,\vec{c}_{n-2}$};
\node at (3.5,-4.8) {$\,\vec{c}_{n-1}$};
\node at (5.5,-4.8) {$\,\vec{c}_{n}$};
\node at (7.5,-4.8) {$\,\vec{c}_{n+1}$};
\node at (9.5,-4.8) {$\,\vec{c}_{n+2}$};

\end{tikzpicture}
\end{equation}

\section{numerical performance}

In this section we provide details on the numerical performance and the convergence of the method with respect to the auxiliary space dimension $\chi_\mathrm{aux}$ and the Trotter time step $\delta t$, using the quench dynamics in the single-impurity Anderson model (SIAM) from Fig.~\ref{fig:hubbard_kondo_quench} (main text) as an example. 

Regarding computational demands, we note that the generation of the semi-group IM via our infinite tensor network is very fast, especially when compared to previous MPS-IM approaches \cite{thoenniss2023Nonequilibrium, ng2023Realtime, guo2024Infinite}. Firstly, the semi-group structure renders memory demands independent of the number of time evolution steps. Note that orthogonalization is not required during the network contraction. The memory bottleneck for generating the IM is thus given by the last SVD in the iTEBD algorithm. This most expensive SVD has to be performed on a matrix of size  $\sim (16\times \chi_\mathrm{aux})\times(16\times \chi_\mathrm{aux})$, where $16=2^4$ is the physical (input-, output) dimension of the IM for a two-state system. 

The numerical demands for a single real-time propagation step with a given IM is identical to finite-MPS IM methods. The state vector that is propagated lives
in the product space of vectorized impurity density matrices and the
auxiliary degrees of freedom. For the SIAM this state has dimensions
$\chi_\mathrm{aux}\times 4 \times 4\times \chi_\mathrm{aux}$, corresponding to
the $\sigma={\uparrow}$ bath, the $\sigma={\uparrow}$ impurity mode, the
$\sigma={\downarrow}$ impurity mode, and the $\sigma={\downarrow}$ bath. To
propagate the state for a single time step we have to multiply this vector with
the semi-group propagators of the two baths. This matrix multiplication can be
performed efficiently on GPUs, which speeds up the propagation significantly in
the case of large auxiliary space dimensions. While finite-time evolution up to time $t$ requires $N=t/\delta t$ such evolution steps, the stationary dynamical regime can be reached more efficiently via Krylov methods, a key advantage of the semi-group structure.

As explained in the main text, we use a fixed relative SVD truncation tolerance in iTEBD which allows to avoid specifying a memory cutoff $N_c$. This is because the action of gates corresponding to very long memory times will fall below the truncation threshold, i.e.~a natural memory cutoff is already specified by the compression tolerance. Note that the specific value of the tolerance does not admit a direct meaning, because the magnitude of singular values depends on the Trotter step $\delta t$, a general property of ``temporal entanglement". However, as demonstrated below, the auxiliary space dimensions (temporal bond dimensions) are comparable for different time step sizes. 

\begin{figure}[h]
\includegraphics[width=0.35\columnwidth]{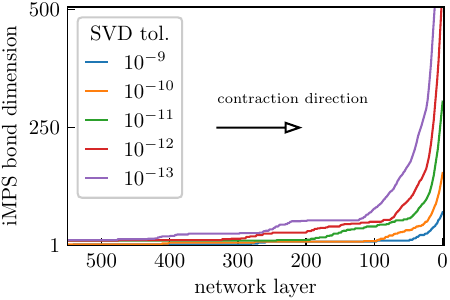}\qquad \qquad 
    \begin{tabular}[b]{cccc}\hline
  SVD tol. & effective $N_c$ & $\,\,\chi_\mathrm{aux}\,\,$ & comp.~time [min] \\ \hline
      $10^{-9}$ & 542  & 71 & 0.87\\
      $10^{-10}$ & 1333 & 154  & 1.34\\
      $10^{-11}$ & 2652  & 305 & 3.46\\
      $10^{-12}$ & 4100 & 556 & 8.09  \\
      $10^{-13}$ & 5564 & 987 & 25.95  \\
\hline
       &  &   \\
       & &  \\

    \end{tabular}

    \caption{Evolution of the iMPS bond dimension during computations of influence matrices for the quench dynamics from Fig.~\ref{fig:hubbard_kondo_quench} (main text), with step size $\Gamma\delta t=0.025$ and different relative SVD cutoff tolerances. The final bond dimension at network layer zero correspond to the auxiliary space dimension $\chi_\mathrm{aux}$. The effective memory cutoff at which the bond dimension starts growing beyond one depends on the specified relative SVD tolerance. The table on the right shows the memory cutoffs as well as exemplary computation times for the contraction on consumer hardware (Apple M4 with linear algebra via OpenBLAS).
    }\label{fig:bonds}
\end{figure}

In Fig.~\ref{fig:bonds} we display the bond dimension growth during network contraction using different SVD tolerances. Significant bond dimension growth occurs only during the last few layers in the network, leading to low computation times. We can also see how the truncation tolerance sets an effective memory cutoff $N_c$, before which the bond dimension remains unity.

\renewcommand{\arraystretch}{1.2}

\begin{figure}[h]
\includegraphics[width=0.35\columnwidth]{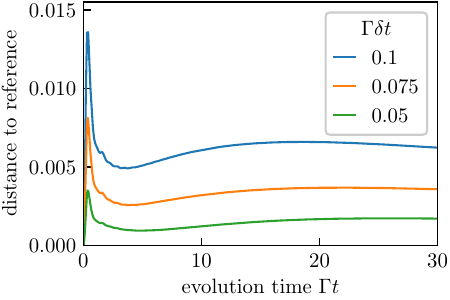}
    \qquad \qquad \includegraphics[width=0.35\columnwidth]{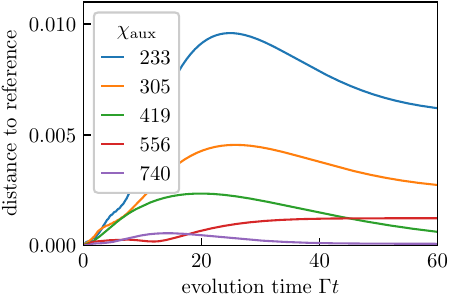}
    \caption{Convergence of the quench dynamics from Fig.~\ref{fig:hubbard_kondo_quench} (main text, left panel). As an error measure we use the Hilbert-Schmidt distance of the impurity density matrix to a reference state $||\rho(t)-\rho_\mathrm{ref}(t)||_2$. For the reference state $\rho_\mathrm{ref}(t)$ we used a simulation with $\Gamma\delta t=0.025$ and $\chi_\mathrm{aux}=987$. Left: Convergence with respect to the Trotter time step $\delta t$. We used large auxiliary space dimensions $\chi_\mathrm{aux}=1086,1185,1257$ for $\Gamma\delta t=0.1,0.075,0.05$ respectively. Right: Convergence for $\Gamma\delta t=0.025$ with respect to $\chi_\mathrm{aux}$ (see also Fig.~\ref{fig:convergence2}). 
    }\label{fig:convergence1}
\end{figure}

In Fig.~\ref{fig:convergence1} the convergence towards a reference state with high $\chi_\mathrm{aux}$ and small $\delta t$ is displayed. The convergence with respect to the Trotter time step $\delta t$ is in line with the expected $O(\delta t^2)$ scaling (see also Ref.~\cite{ng2023Realtime}). We moreover find consistent convergence with respect to the auxiliary space dimension $\chi_\mathrm{aux}$. 

\begin{figure}
    \includegraphics[width=0.35\columnwidth]{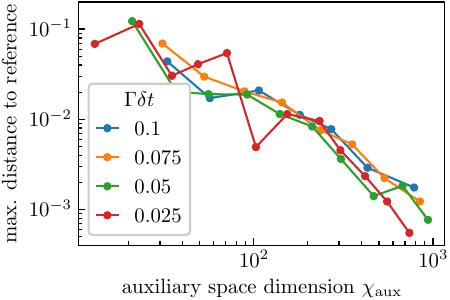}
    \qquad \qquad \includegraphics[width=0.35\columnwidth]{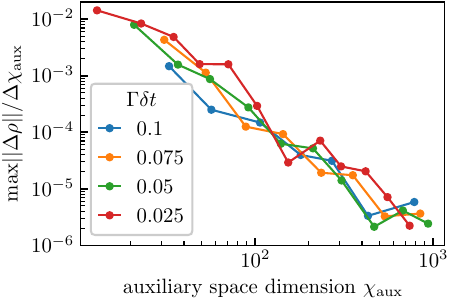}
    \caption{Convergence of the quench dynamics from Fig.~\ref{fig:hubbard_kondo_quench} (main text, left panel) with respect to $\chi_\mathrm{aux}$ for different Trotter time steps $\delta t$. Left: Convergence of the dynamics. As an error measure we use the maximum Hilbert-Schmidt distance of the impurity density matrix to a reference state over the full evolution time $\mathrm{max}_t\left(||\rho(t)-\rho_\mathrm{ref}(t)||_2\right)$. For the reference states $\rho_\mathrm{ref}(t)$ we used $\chi_\mathrm{aux}=1086,1185,1257,987$ for $\Gamma\delta t=0.1,0.075,0.05,0.025$ respectively. Left: Maximum distance to the simulation with next-larger bond dimension, as in Eq.~\eqref{eq:distance}. 
    }\label{fig:convergence2}
\end{figure}

In Fig.~\ref{fig:convergence2} the convergence with respect to the bond dimension $\chi_\mathrm{aux}$ for different Trotter step sizes $\delta t$ is displayed. The error appears to be independent of $\delta t$.
For estimating the convergence speed we determine numerically the scaling exponent $r$ of the error order $O(1/(\chi_\mathrm{aux})^r)$. In order to extract the exponent without knowledge of the exact solution, we consider the maximum distance between simulations with respect to $\chi_\mathrm{aux}$
\begin{equation}\label{eq:distance}
\frac{\mathrm{max}_t\left(||\rho_2(t)-\rho_1(t)||_2\right)}{|\chi_\mathrm{aux,2}-\chi_\mathrm{aux,1}|}\sim O\left(1/(\chi_\mathrm{aux,1})^{(r+1)}\right).
\end{equation}
As can be seen in Fig.~\ref{fig:convergence2} (right panel), we indeed find algebraic scaling and we extract $r\approx 1.2$ independent of the Trotter step size.

\end{document}